# Strong Raman Optical Activity and Chiral Phonons in Chiral Hybrid Organic-Inorganic Perovskites

*Evan W. Muller, Aleksey Ruditskiy, Jie Jiang, Thuc T. Mai, Katherine Burzynski, Ruth Pachter, Michael F. Durstock, W. Joshua Kennedy\*, Rahul Rao\**

Evan W. Muller, Aleksey Ruditskiy, Jie Jiang, Thuc T. Mai

AeroVironment Inc., Dayton, OH, 45433 United States

Katherine Burzynski, Ruth Pachter, Michael F. Durstock, W. Joshua Kennedy, Rahul Rao

Materials and Manufacturing Directorate, Air Force Research Laboratory, Wright-Patterson Air Force Base, OH, United States

E-mail: william.kennedy.21@us.af.mil, rahul.rao.2@us.af.mil

**Abstract**

Hybrid organic-inorganic perovskites with chiral organic cations are very interesting for optoelectronic applications because of their intrinsically chiral light-matter interactions. Chiral distortions in these materials lead to circular dichroism, circular birefringence, and circularly polarized luminescence in the band transitions of the inorganic sublattice. Raman-active vibrational modes in these crystals are governed by crystal symmetry and therefore are also strongly impacted by the nature and magnitude of the chiral distortions. Here, we report low-frequency Raman modes that are sensitive to circularly polarized excitation in chiral hybrid organic-inorganic perovskites (CHOIPs) across a wide range of structures and compositions. The circularly polarized Raman spectra from enantiomers of CHOIP single crystals exhibit sharp modes below 150 cm$^{-1}$, corresponding to vibrations of the lead iodide octahedra. These modes exhibit strong differences in intensities (Raman optical activity, ROA) depending on the handedness of the excitation, with high degree of polarization for several modes. Calculations reveal the presence of several chiral phonon modes with opposite phonon angular momenta. The strong ROA and the chiral phonon modes are a direct consequence of chirality transfer from the chiral organic linker to the lead iodide octahedra in the CHOIP structure, resulting in a strong chiroptical response in the phonon modes.

**Introduction**

Hybrid organic metal halides make up a large family of materials that have garnered a broad interest for optoelectronic applications because of their ease of synthesis and processing, strong optical cross-sections, and compositionally tunable spectral responses.[1] Despite the wide range of crystal structures that emerge, these materials are generally referred to as hybrid organic-inorganic perovskites (HOIPs) due to the prevalence of octahedral coordination in the metal-halide sublattice.[1,2] When non-centrosymmetric chiral organic cations are incorporated within the metal halides, they can impart a chiral distortion onto the metal halide octahedra, altering the lengths and angles of some of the metal halide bonds and imparting a twist to the sublattice octahedra.[3–5] For some chiral hybrid organic-inorganic perovskites (CHOIPs), this results in a non-centrosymmetric space group, while in other cases the crystal is centrosymmetric but the lead-halide sublattice has a non-centrosymmetric point group.[6] We note that, in this manuscript, we only consider CHOIPs with long-range translational symmetry rather than the myriad heterostructures comprising the broader set of hybrid materials such as Ruddlesden-Popper phases where surface distortions are distinct from the equilibrium bulk structure.[7–10]

Many CHOIPs exhibit strong polarization-dependent optical properties including circular birefringence, circular dichroism (CD), and circularly polarized luminescence (CPL).[7,11] The strong CD in the optical transitions associated with the metal halide sublattice has been leveraged for polarization sensitive photodetection and for CPL.[12–15] In general, the magnitude of chiroptical response in interband transitions in CHOIPs is related to the degree of chiral distortion imparted to the octahedral metal halide sublattice. For example, distortions of the lengths and angles in the metal halide bonds can be used in methylbenzylammonium (MBA) lead bromides doped with methylbutylammonium (MeBA) to explain subtle differences in the CD spectra.[4] Additionally, chirality can also impact Rashba-Dresselhaus spin-orbit coupling and spin splitting,[5,16] we well as structural metrics such as metal halide bond length distortion $\Delta d$, bond angle variance $\sigma^2$, and disparity in adjacent bond angles $\Delta\beta$. These distortions should have a direct impact on phonon modes, providing a direct method to study chiral light-matter interactions in CHOIPs.

Low frequency (< 100 cm$^{-1}$) Raman scattering spectra from HOIPS are dominated by modes pertaining to the metal halide bonds and their interactions with the organic cations, while the higher frequency region (> 700 cm$^{-1}$) corresponds to the bending, stretching and rocking modes of the organic moieties.[17,18] Previous studies on organic (CH$_3$NH$_3$PbBr$_3$) and inorganic (CsPbBr$_3$) perovskites have reported broadened Raman peaks at room temperature that were attributed to anharmonic polar fluctuations of the lead halide octahedra.[19,20] Temperature-dependent Raman studies of various HOIPs have also revealed anomalous changes in peak frequencies and linewidths, indicative of structural phase transitions in the materials.[21–23]

Considering that the Raman modes are sensitive to small variations in the metal halide bond lengths, they should also be influenced by the chirality transferred from the chiral organic cations to the metal halide octahedra. Chiroptical spectroscopy methods to measure vibrational modes include Raman optical activity (ROA), vibrational CD and terahertz CD. Of these methods, ROA is a powerful technique to study the structure of chiral molecules by measuring the difference of Raman scattering intensities under right circularly polarized (RCP) and left circularly polarized (LCP) excitations.[24–26] Circularly polarized Raman spectroscopy can also be used to directly probe the chirality of phonon modes, *i.e.* modes where an atom or a group of atoms in a solid undergo rotational motion perpendicular to the direction of propagation of the vibration.[27] These so-called chiral phonons exhibit angular momenta, which manifest as a slight (up to a few cm$^{-1}$) splitting of degenerate mode dispersions near the center of the Brillouin zone, and hence measurable by employing RCP and LCP excitations. These modes may also exhibit ROA. Recent circularly polarized Raman measurements on chiral and achiral materials such as quartz, tellurium, HgS, and two-dimensional ReSe$_2$, ReS$_2$ and AgCrP$_2$Se$_6$ have revealed the presence of chiral phonons, which, depending on the material may exhibit both splitting of mode frequencies or changes in relative intensities.[28–34] In the case of CHOIPs, a recent computational study predicted low-frequency chiral phonons corresponding to vibrations of the lead iodide framework in two-dimensional perovskites such as MBA$_2$PbI$_4$.[35] However, to our knowledge, there are no published reports of circularly polarized Raman measurements on CHOIPs.

In this work, we have synthesized and characterized a series of lead iodide-based CHOIPs, selecting a group of chiral organic cations with varying sizes and number of cyclic structures including several naturally occurring amino acids. Structural characterization using X-ray diffraction (XRD) and CD spectroscopy revealed the transfer of chirality from the organic cations to the lead iodide octahedra. This transfer results in low-frequency Raman-active modes in the inorganic lead iodide octahedra that can be selectively excited by circularly polarized light in enantiomers of CHOIP single crystals. We observed several sharp (widths as low as 4 cm$^{-1}$) and intense Raman modes at room temperature that persist upon heating up to 400 K. ROA measurements conducted with sub-band gap excitation (785 nm or 1.58 eV) show that several of the Raman modes exhibit clear differences in intensities with RCP and LCP excitations, *i.e.*, they exhibit ROA. We modeled the normal modes of the CHOIP crystals and show that many of the low energy vibrations comprise non-colinear movements of the constituent atoms, suggesting that the phonons themselves are chiral. Our experimental observations were confirmed by phonon angular momentum calculations in (S-NEA)PbI$_3$, which revealed small but finite splitting of several degenerate phonon branches at the center of the Brillouin zone. The high scattering cross-section and strong polarization sensitivity provide a deeper understanding of the underlying mechanism for chiral light-matter interactions in CHOIPs.

**Results and Discussion**

The chemical structures of the organic cation enantiomers are shown in **Figure 1a**. The lead iodide-based CHOIPs were crystallized solvothermally from solutions of the organic amines or amino acids and lead iodide salts dissolved in hydroiodic acid. The resulting crystals (**Figure S1** shows optical images of some of the CHOIP single crystals) were analyzed with X-ray diffraction (XRD) to determine their structure. The patterns confirm the chiral space groups of the synthesized CHOIPs. Moreover, we used the XRD data to calculate chiroptical metrics and quantify the chiral distortions on the metal-halide octahedra imparted by the organic cations (see **Table S1** and accompanying discussion).

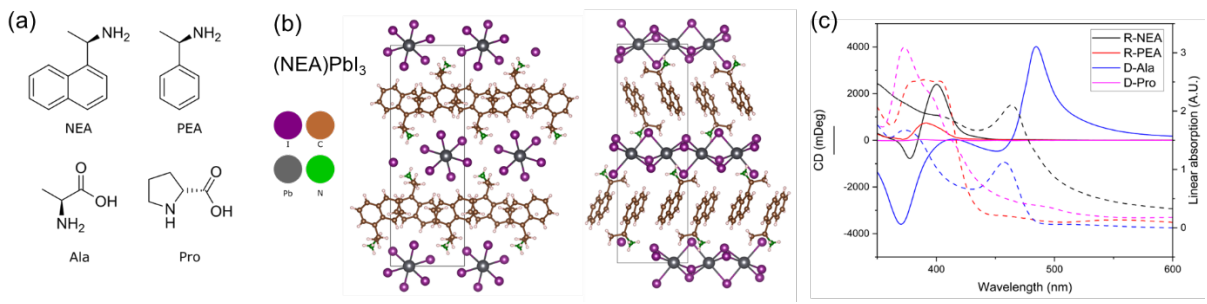

**Figure 1**. Structural and spectroscopic data for CHOIPs. a): R/D enantiomer structures of chiral organic molecules. b) Top view (left) and side view (right) of the structure of (R-NEA)PbI$_3$. c) UV-Vis / ECD spectra of CHOIPs. The filled lines correspond to ECD (left axis) and dotted line corresponds to UV-Vis (right axis).

For each CHOIP composition we fabricated polycrystalline thin films by spin coating solutions on glass as described in the Experimental section. The resulting thicknesses of the polycrystalline films were 400 ± 70 nm across all samples as measured by profilometry. We measured the absorption and CD spectra in the UV-visible wavelength range from several films of each material, taking the average of multiple spectra obtained in different film orientations to correct for linear birefringence and to account for differences in film thicknesses generated during the spin-coating process. **Figure 1c** shows the CD (left axis) and UV-Vis (right axis) absorption spectra for four CHOIPs – (R-NEA)PbI$_3$, (R-PEA)PbI$_3$, (D-alanine)PbI$_3$ and (D-proline)PbI$_3$. All the CHOIPs show semiconductor-like optical absorption spectra (dashed lines in **Figure 1c**) with a strong absorption edge or excitonic peak near the band gap; the absorption peak corresponds to excitons associated with interband transitions in the PbI$_3$ sublattice. The largest deviation in transition energy is observed in the (D-alanine)PbI$_3$, in which the reduced quantum confinement of the wider quasi-1D sublattice leads to a decrease in the band gap relative to the other structures. We note that some of the absorption spectra contain a small shoulder near 500 nm arising from the presence of PbI$_2$ precipitates in the spin cast films that arise from small amounts of perovskite decomposition during thermal annealing of the films.

Each CHOIP exhibits CD (solid traces in **Figure 1c**) near the inter-band transition of the PbI$_3$ sublattice, far below the HOMO-LUMO transitions of the constituent organic. The CD spectra all exhibit a positive Cotton effect and the anisotropy factor $g_{CD}$ is comparable to literature reports of similar materials. The appearance of strong circular dichroism is direct evidence for the chiral distortions present in our CHOIPs and motivates an investigation of the polarized Raman scattering in these materials.

We measured low-frequency circularly polarized Raman spectra from multiple single crystals of several CHOIP enantiomers with an excitation wavelength (785 nm, 1.58 eV) well below the bandgap of the CHOIPs. The sub-gap excitation ensured that the observed Raman peak intensities were not affected by resonance conditions.[36] Right and left circularly polarized excitation (RCP and LCP, respectively) were produced using a combination of quarter and half-waveplates. In our backscattering configuration, the scattered light passes through the same waveplates as the excitation and its polarization can be switched using an additional half waveplate prior to the light entering the spectrometer. In this way we can measure spectra in co- and cross-circularly polarized configurations (RR, LL, RL and LR; **Figure S2** shows an image of our optical layout).

An example set of cross-circularly polarized Raman spectra from (R/S-NEA)PBI$_3$ single crystals are shown in **Figure 2**, which plots the low-frequency Stokes and anti-Stokes regions (-180 to 180 cm$^{-1}$). The spectra in **Figure 2** were collected at room temperature and exhibit sharp peaks, with the lowest frequency peak appearing around 10 cm$^{-1}$ in the LR spectra from both R- and (S-NEA)PbI$_3$. The room temperature linewidths (full width at half maximum intensities) of some of these low-frequency peaks are as low as 4 cm$^{-1}$. Such narrow peak widths in the Raman spectra from hybrid organic-inorganic perovskites are unusual at room temperature. Previous Raman measurements on achiral hybrid organic-inorganic as well as all-organic lead halide perovskites revealed significantly broadened low-frequency peaks at room temperature and were attributed to anharmonic polar fluctuations in the lead halide lattice.[19] Our low peak linewidths indicate a significant reduction in the room temperature polar fluctuations in the PbI$_3$ octahedra, which we attribute to the high degree of structural rigidity from the organic chiral linkers. Temperature-dependent Raman measurements confirmed this hypothesis. Taking (R/S-NEA)PbI$_3$ as an exemplar material, we measured Raman spectra from 100 – 450 K. Upon cooling, the Raman peaks in the spectra (shown in **Figure S3**) sharpened as expected, with blue shifted frequencies due to thermal contraction. Upon heating, we did not see any significant changes in the spectra up to 400 K, other than the usual thermally induced peak broadening and redshifts. At the highest temperature of 450 K, we observed severe broadening and diminishing of the Raman peaks that can be attributed to thermal degradation. But the persistence of all the Raman peaks between 100 – 400 K indicates a lack of any phase transitions in the quasi-1D and quasi-2D CHOIPs, in contrast to those commonly seen in the 3D hybrid organic-inorganic perovskites.[2] This structural stability extends the operating range for the CHOIPs in optoelectronic and phononic applications.

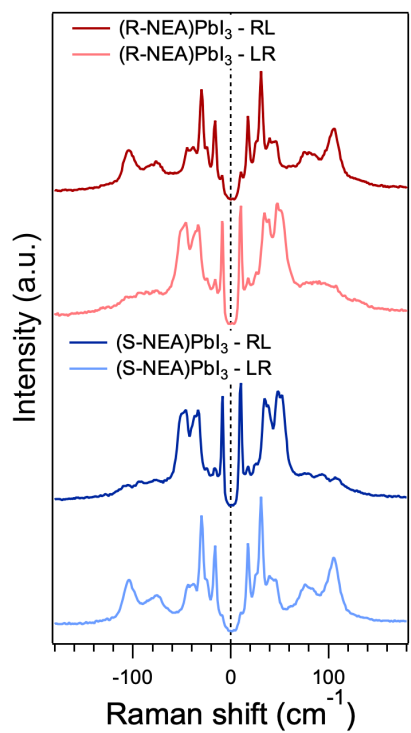

**Figure 2.** Cross-circularly polarized (RL and LR) Raman spectra in the low-frequency Stokes and anti-Stokes regions from (R/S-NEA)PBI$_3$.

Lorentzian peak fitting analysis revealed 10 peaks (**Figure S4**) between 0 – 150 cm$^{-1}$ in (R-) and (S-NEA)PbI$_3$ and 12 peaks in the unpolarized Raman spectra collected at room temperature. **Table 1** lists the peak frequencies corresponding to the cross-circularly polarized (RL and LR) and unpolarized spectra from (S-NEA)PbI$_3$. To gain further insights into the nature of these vibrational modes, we performed density functional theory (DFT) calculations (details in the Experimental section) of the Raman spectrum of (S-NEA)PbI$_3$, specifically for understanding the low frequency modes related to the PbI$_3$ octahedra. Calculations were performed at 0 K with a broadening parameter of 3 cm$^{-1}$ and for simplification, unpolarized light was considered. The calculations reproduced all 12 experimentally observed peaks (**Figure S5** and **Table 1**) in the unpolarized (S-NEA)PbI$_3$ Raman spectrum, demonstrating only small blueshifted peak frequencies (between 0.7 - 7.7 cm$^{-1}$). We note that the total number of phonon modes at the Γ point at 0 K is much higher than the number of phonon modes observed experimentally at room temperature. Of the 369 Γ point at 0 K Raman-active modes, 61 of the modes in the lower frequency range (0-100 cm$^{-1}$) correspond to vibrations of the PbI$_3$ octahedra. Not all these modes can be observed at room temperature because many of them have intensities that are too weak and/or frequencies that are too close to observe in the measured Raman spectrum.

Table 1. Measured (RL, LR and unpolarized) and calculated Raman peak frequencies for (S-NEA)PbI$_3$. The chiral modes (peaks 1, 2, 4 and 7) are highlighted in bold font.

| Peak | Measured, RL (cm$^{-1}$) | Measured, LR (cm$^{-1}$) | Measured, unpolarized (cm$^{-1}$) | Calculated (cm$^{-1}$) |
|---|---|---|---|---|
| 1 | 10 | 10.1 | 10 | **10.7** |
| 2 | 17.3 | 17.7 | 17.4 | **20.1** |
| 3 | 25.4 | 25.4 | 25.2 | 32.1 |
| 4 | 34.1 | 31.1 | 33.9 | **36** |
| 5 | 38.9 | 39.3 | 39.7 | 39 |
| 6 | 47.6 | 42.8 | 47.5 | 50.9 |
| 7 | 52.7 | 46.6 | 53 | 53.1 |
| 8 | 77.3 | 77.7 | 61.4 | **58.5** |
| 9 | 94 | 101.1 | 75.4 | 83.7 |
| 10 | 108.8 | 106.5 | 90.8 | 99.8 |
| 11 | | | 107.5 | 105.8 |
| 12 | | | 131.5 | 136.4 |

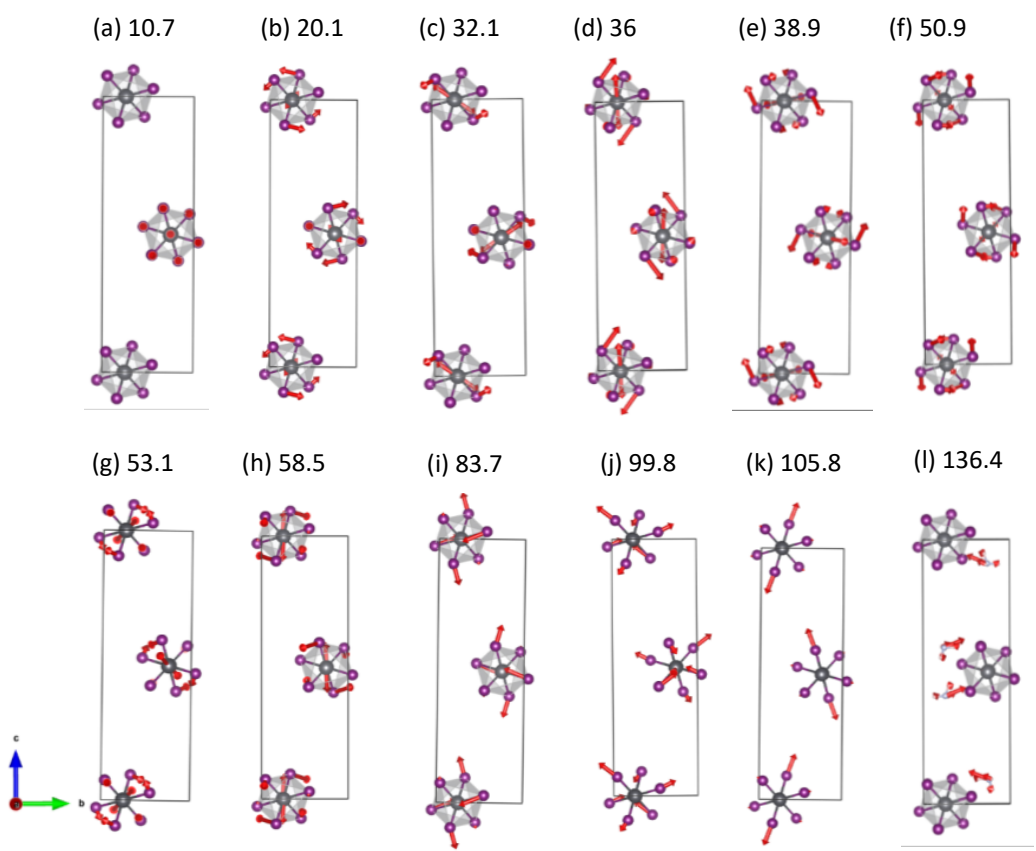

**Figure 3.** a) – i) Top view of phonon eigenvectors shown along the PbI$_3$ octahedra for the nine calculated Raman peaks in (S-NEA)PbI$_3$. The 1-D axis of the PbI$_3$ chains is perpendicular to the plane of the page. The Pb and I atoms are shown in grey and purple, respectively. The red arrows indicate the directions of the eigenvectors. In i), j), and k) the organic cation vibration is not shown for clarity.

The phonon eigenvectors for the 12 calculated (S-NEA)PbI$_3$ Raman modes are shown in **Figures 3 and S6** in two different views (from the top and along the *a* axis, respectively). In general, the modes up to 60 cm$^{-1}$ (**Figures 3a-3h**) correspond mainly to vibrations of the PbI$_3$ octahedra, with small contributions from the organic cations, while the highest frequency mode at 136.4 cm$^{-1}$, **Figure 3l**) involve organic cation vibrations by NH$_3$. The modes from 60 to 110 cm$^{-1}$ (**Figures 3i-3k**) correspond to both the PbI$_3$ and organic cation vibrations.

In addition to sharp low-frequency peaks, the other striking feature observed in the Raman spectra in **Figure 2** is that, for a given enantiomer, the relative intensities of the low-frequency peaks are markedly different in the RL and LR spectra, with some peaks appearing/disappearing completely while others exhibiting variations in intensities. In fact, the RL spectrum from (R-NEA)PbI$_3$ is identical to the LR spectrum from (S-NEA)PbI$_3$, while the LR spectrum from (R-NEA)PbI$_3$ is identical to the RL spectrum from (S-NEA)PbI$_3$. We note that similar spectra were obtained with the co-circularly polarized configuration (i.e. RR and LL, **Figure S7**), suggesting that the large variation in peak intensities between the two sets of spectra are due to preferential absorption of RCP or LCP excitation by the two CHOIP enantiomers. While the difference in intensities was the strongest in the low-frequency region, we confirmed that the differential scattering of RCP and LCP excitation was also reflected in the high-frequency spectral region, i.e. the region dominated by the Raman peaks corresponding to the organic linkers. **Figure S8** shows an extended range (0-1650 cm$^{-1}$) of the circularly polarized Raman spectra from (R/S-NEA)PbI$_3$ with a magnified view of the 400 – 1600 cm$^{-1}$ spectral region. Several weak peaks can be observed with higher intensities for RCP or LCP excitations. These high-frequency Raman modes correspond to the common fingerprint region for organic molecules, and comprise of stretching and bending modes involving carbon, hydrogen and nitrogen atoms.

The difference in peak intensities with RCP and LCP excitation can be clearly seen in the Raman optical activity (ROA), which we define as the difference between the intensities of the RCP and LCP excited spectra. In typical ROA experiments, either the incident or scattered light is circularly polarized, while the other is left un-polarized.[37] Moreover, the incident or scattered circularly polarized light in an ROA measurement is modulated continuously similar to a CD spectrometer.[24,38] Our optical setup allows for the static collection of Raman spectra in co- or cross-circularly polarized Raman scattering. However, considering the preferential absorption of RCP and LCP excitations by the enantiomers of the CHOIPs (**Figure 2**), the difference between the intensities of the RCP and LCP excited spectra do provide a measure of ROA ($I_{RL} - I_{LR}$ or $I_{RR} - I_{LL}$, labeled hereafter as $I_R - I_L$ for brevity) from the CHOIPs.

The ROA spectra from (R/S-NEA)PbI$_3$ are shown in the top panel of **Figure 4a**. We see very strong bisignate ROA bands across the whole low-frequency range, from 10 – 150 cm$^{-1}$. In contrast, the ROA from the high frequency region (**Figure S8**) was very poor. The presence of both positive and negative bands in **Figure 4a** indicates preferential absorption of RCP and LCP excitation, which can directly be attributed to the complex chiral arrangement of the PbI$_3$ octahedra. We also see similar ROA with other CHOIPs. **Figure 4a** also shows ROA from (R/S-PEA)PbI$_3$, (D/L-alanine)PbI$_3$, and (D/L-proline)PbI$_3$. Similar to (R/S-NEA)PbI$_3$, the other three CHOIPs also exhibit strong ROA across the whole low frequency range (10 – 150 cm$^{-1}$). Interestingly, the maxima/minima appear at distinctly different frequencies in the CHOIPs.

**Figure 4a** also shows that we observe ROA from chiral amino acid-based perovskites, which are structurally much more complex than the NEA- or PEA-based perovskites.

The strength of the chiroptical activity can also be seen in the circular intensity difference (CID, also known as the degree of circular polarization), which is given by $(I_R - I_L)/(I_R + I_L)$. Here $I_R + I_L$ represents unpolarized spectra. The CID for (R/S-NEA)PbI$_3$, (R/S-PEA)PbI$_3$, (D/L-alanine)PbI$_3$ and (D/L-proline)PbI$_3$ are shown in **Figure 4b**. Overall, we see very high values, the highest being 0.73 at 9 cm$^{-1}$ for (S-NEA)PbI$_3$. These CID values are very high compared to other organic chiral materials, which typically exhibit CIDs a few orders of magnitude lower (~$10^{-3}$)[37] but they are closer to those measured recently in two dimensional materials such as ReS$_2$, ReSe$_2$ and AgCrP$_2$Se$_6$.[31–34]

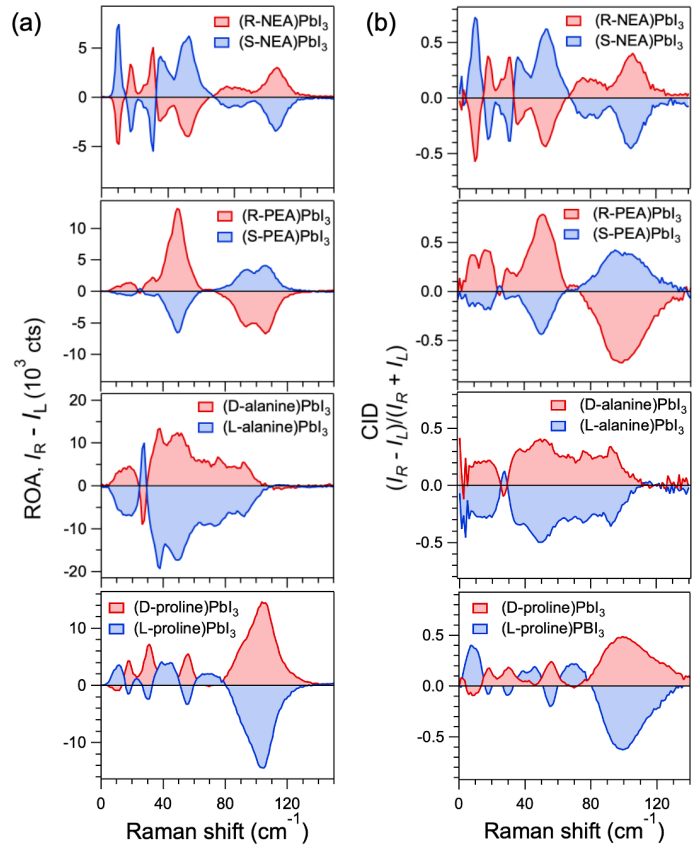

**Figure 4.** a) ROA and b) CID spectra from (R/S-NEA)PbI$_3$, (R/S-PEA)PbI$_3$, (D/L-alanine)PbI$_3$ and (D/L-proline)PbI$_3$.

The observation of strong ROA raises the question of whether the CHOIPs also host chiral phonons. To answer this, we used DFT to calculate the phonon angular momentum (PAM) at 0 K in (S-NEA)PbI$_3$. The PAM expression projects the phonon eigenvector onto an angular momentum operator, where for a phonon eigenvector $u_{q,\mu}$ the PAM ($J_{q,\mu}^\alpha$) ($\alpha = x, y, z$) is given by $J_{q,\mu}^\alpha = \hbar u_{q,\mu}^+ M_\alpha u_{q,\mu}$ in terms of the matrix product $M_\alpha = I_{N\times N} \otimes \begin{pmatrix} 0 & -i\varepsilon_{\alpha\beta\gamma} \\ -i\varepsilon_{\alpha\beta\gamma} & 0 \end{pmatrix}$. Here N is the number of atoms in a unit cell and $\varepsilon_{\alpha\beta\gamma}$ the Levi-Civita epsilon tensor defining the cross-product structure in 3D.[39–41] (S-NEA)PbI$_3$ belongs to the P2$_1$2$_1$2$_1$ space group, and chiral phonons were previously calculated for (S-MBA)$_2$PbI$_4$, which belongs to the same space group.[35] The calculated PAM values (**Figure 5**) for (S-NEA)PbI$_3$ clearly indicate that chiral phonons are present in the lower frequency range below 150 cm$^1$, with opposite signs along Γ-X vs ΓX (Γ-Y vs ΓY or Γ-Z vs ΓZ) due to time reversal symmetry. The PAM values for an extended frequency range (up to 220 cm$^{-1}$) are shown in **Figure S9**.

Specifically, the color-coded branches in **Figure 5** show opposite PAM values for some phonon modes near the Γ point as well as towards the zone boundaries. Our results are consistent with previous calculations for (S-MBA)$_2$PbI$_4$.[35] The 12 calculated Raman peaks (**Table 1**) are indicated by blue dots in **Figure 5**. Chiral Raman modes are observed, for example for the calculated peak at 10.7 cm$^{-1}$, with a large angular momentum $J_z$ along the Γ-Z direction of the Brillouin zone. The phonon mode at 58.5 cm$^{-1}$ has a large negative $J_y$ along the Γ-Y direction.

The phonon dispersion and PAM values for two of the modes are shown **Figure 6**. The mode near 10.7 cm$^{-1}$ comprises two nearly degenerate modes at 10.68 and 10.69 cm$^{-1}$ with weaker and stronger intensities, respectively (upper panel of **Figure 6a**), which are further split along the ΓZ direction at $k$=Z/10 with frequencies of 10.63 and 10.67 cm$^{-1}$, respectively. The phonons at Z/10 demonstrate circular vibration of the atoms (see animations in Supplementary Movies **M1**). **Figure 6b** shows the phonon eigenvectors for the mode at 10.67 cm$^{-1}$. Each atom rotates around its center along $z$ axis, giving rise to large (near -1.0) negative $J_z$. The mode at 10.63 cm$^{-1}$ has a similar phonon eigenvector but with anticlockwise rotation, giving rise to large (near 1.0) positive $J_z$. Animations of the modes can be seen in Supplementary Movie **M2**. Note that the lack in inversion symmetry in the P2$_1$2$_1$2$_1$ space group allows degenerate Γ-point phonons to carry angular momentum, such that the nearly degenerate Γ mode near 10.7 cm$^{-1}$ comprises a non-circular mode vibrating linearly along the *a* axis (Figures **3a** and **S6a**) and another along the *b* axis (not shown). The two linear vibrations combine to form complex eigenvectors that vibrate circularly in the *ab* plane (**Figure 6b**).

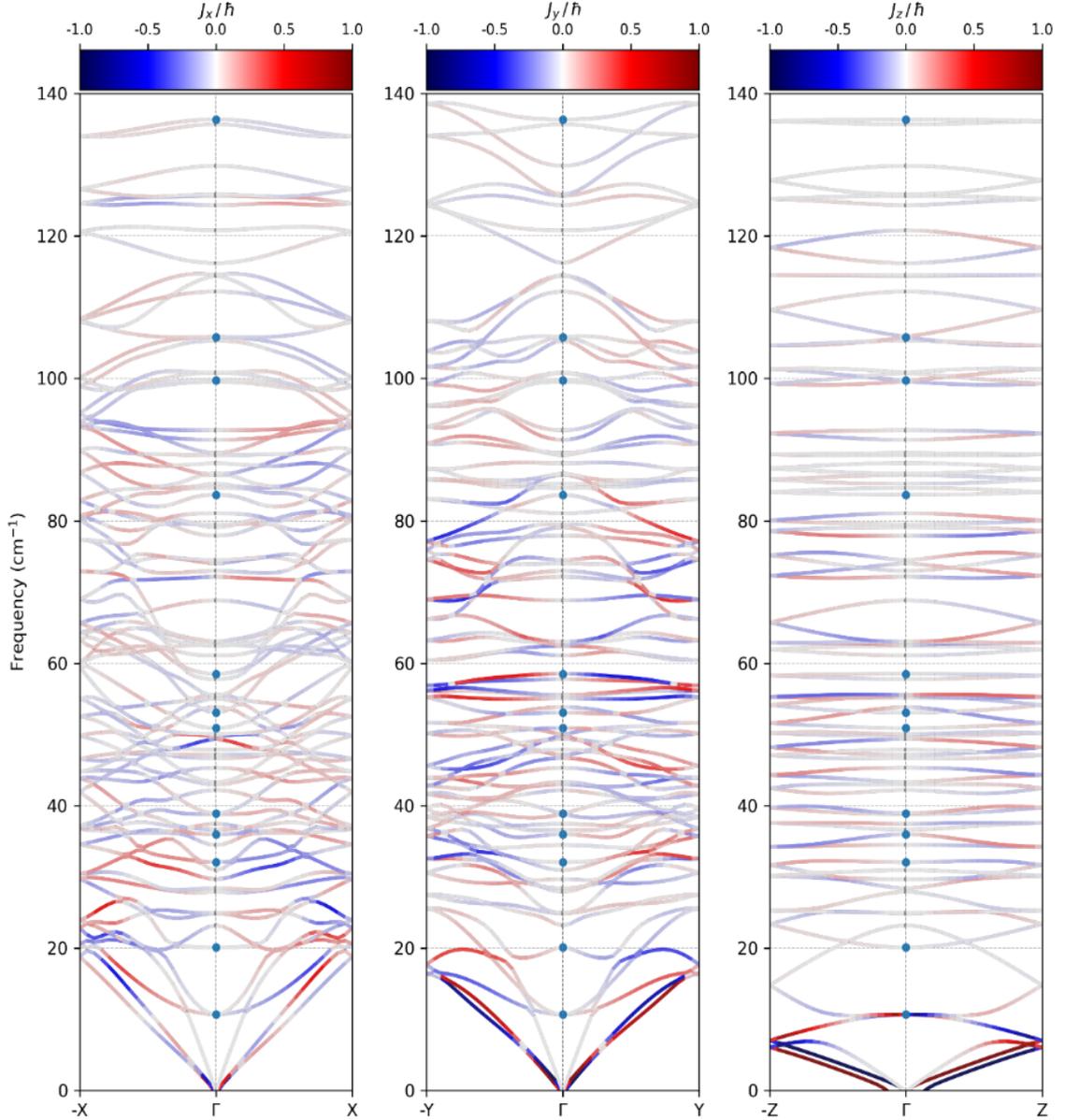

**Figure 5.** PAM values in (S-NEA)PbI$_3$ for phonons between 0 – 140 cm$^{-1}$ along the ΓX, ΓY, and ΓZ directions of the Brillouin zone. The experimentally measured Raman modes are shown as dots at the Γ point.

The phonon dispersion and PAM values for the mode near 58.5 cm$^{-1}$ are shown in the lower panel of **Figure 6a**. The phonon branch along ΓY is nearly flat and the phonon at Y/10 has a negative $J_y$. The phonon vibration for the mode at Y/10 shows a different circular vibration (animations in Supplementary Movies **M3**), as depicted in **Figure 6c**. The vibrations in the Pb$_2$I$_6$ units are dominated by two Pb and two I atoms. The Pb atoms rotate clockwise around their center along the $b$ axis, giving rise to negative $J_y$. The I atoms rotate around their center along its bond to Pb (Pb-I bond). Since the two I atoms in the upper $Pb_2I_6$ unit rotate clockwise and the two I atoms in the lower $Pb_2I_6$ rotate anticlockwise, their contribution to $J_z$ is cancelled, leaving net angular momentum contribution to $J_y$. However, since the I atoms do not fully contribute to $J_y$, the magnitude of the total $J_y$ is smaller than 1.0. Interestingly, the four phonon modes at the CID peaks highlighted in bold in Table 1 are found to have finite PAM in **Figure 5**, specifically, the mode at 10.7 cm$^{-1}$ - $J_z$, the mode at 20.1 cm$^{-1}$ - $J_y$, the mode at 36 cm$^{-1}$ - $J_x$, $J_y$, and $J_z$, and the mode at 53.1 cm$^{-1}$ - $J_y$.

Unfortunately, the splitting of frequencies between the modes with opposite PAM (e.g. 10.67 and 10.63 cm$^{-1}$) is too small for us to measure experimentally. Moreover, considering the complex phonon dispersion of the CHOIPs, it is difficult to find the exact correspondence between the chiral phonons measured experimentally, the ROA intensities and the calculated modes. Nevertheless, our measurements and calculations demonstrate that the CHOIPs host chiral phonon modes, and that the chiral distortions in CHOIPs lead to strong polarization response in Raman scattering.

With the wide variety of structures possible in the CHOIPs, the manipulation and study of chiral phonons offers exciting opportunities for emerging applications in phononics. In addition, control over electron-(chiral) phonon coupling can afford tunability over circularly polarized luminescence for optoelectronics applications as well as the potential for development of chiral Raman lasers using these materials.

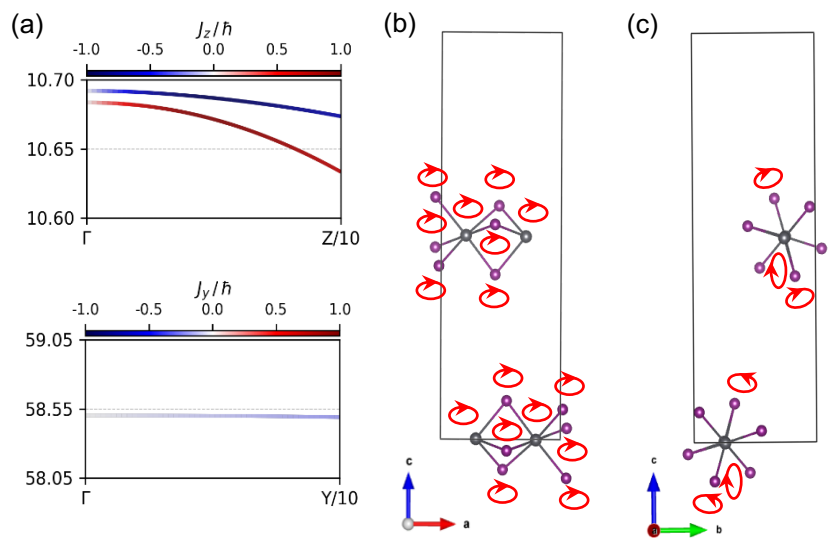

**Figure 6.** a) Phonon dispersion and PAM for phonons near 10.7 and 58.5 cm$^{-1}$. b) schematic phonon vibration for the mode at 10.67 cm$^{-1}$ at $k$ =Z/10, and c) schematic phonon vibration for the mode at 58.5 cm$^{-1}$ at $k$ =Y/10.

**Conclusions**

The low frequency (below 150 cm$^{-1}$) Raman modes in CHOIPs are strongly polarization dependent, exhibiting large variations in intensities upon circularly polarized excitation. These modes also exhibit opposite phonon angular momenta due to their chiral nature. The universality of the circularly polarized Raman scattering in CHOIPs across a wide range of organic cations and crystal structures is a striking and surprising phenomenon that offers a new approach to measuring chirality transfer in the CHOIPs, as well as for tailoring polarized light-matter interactions in these materials. Vibrational modes are relatively straightforward to measure and calculate, and the non-colinear motions of individual atoms in the vibrational modes are a strong indicator of Raman optical activity. Future studies could systematically explore the relationship between specific chiral distortions and the polarization anisotropy. The strongly selective scattering processes may allow for polarization control of specific vibrational modes as well as the development of circular polarization filters, modulators, and Raman lasers.

**4. Experimental**

*Crystal synthesis*

(R)-(+)-1-(1-naphthyl)ethylamine, (S)-(-)-1-(1naphthyl)ethylamine, (R)-(+)-1-phenylethylamine, and (S)(-)-1-phenylethylamine were purchased from Chem-Impex Intl. D-alanine, L-alanine, D-proline, L-proline, anhydrous N,N-dimethylformamide, and hydriodic acid (47%) were purchased from Sigma Aldrich. Lead (II) iodide was purchased from Fisher Scientific. All chemical reagents were used as-received without further purification.

Crystals of all CHOIPs apart from the amino acid perovskites were grown in a similar manner. An equimolar amount of the chiral amine and lead (II) iodide were dissolved in HI (12 mL) and heated to 120 °C followed by slow cooling (5 ◦C / hour) to achieve crystal growth. The crystals were washed with CH$_2$Cl$_2$ to remove residual HI, then sheltered from light. Below are the specific synthetic details for synthesizing each CHOIP:

R/S−NEA−PbI$_3$: R/S-NEA (0.34 g, 2 mmol) and PbI$_2$ (0.92 g, 2 mmol) resulting in yellow, flake to prismatic-shaped crystals (0.3g, 55%).

R/S−PEA−PbI$_3$: R/S-PEA (0.24 g, 2 mmol) and PbI$_2$ (0.92 g, 2 mmol) resulting in highly anisotropic cm-scale yellow, prismatic crystals (0.68g, 66%).

D/L−Alanine−PbI$_3$: D/L-alanine (0.18 g, 2 mmol) and PbI$_2$ (1 g, 2.2 mmol) were dissolved in HI (8 mL) and placed in a refrigerator for two days to allow for growth of the perovskite crystals. The resulting yellow-orange, rectangular crystals were removed from the HI solution and washed with diethyl ether (0.42 g, 44%).

D/L-Proline-PbI$_3$: D/L-proline (0.32 g, 2 mmol) and PbI$_2$ (1g, 2.2 mmol) were dissolved in HI (8 mL) and placed in a refrigerator for two days to allow for growth of the perovskite crystals. The resulting yellow rectangular crystals were removed from the HI solution and washed with diethyl ether (0.28 g, 34%).

*Thin-film fabrication methodology*

Perovskite solutions were prepared by combining the chiral amine salt and lead (II) iodide in 0.75:1 molar ratio in DMF along with one weight percent additive of methylammonium iodide to achieve a 1 M solution. The solutions were sonicated for 10 minutes followed by filtration through a 0.22 μm PVDF syringe filter to yield the final spin-coating solution. Glass substrates with dimensions 1" x 1" were prepared by sonication in a sequence of mucasol/ DI water, DI water, methanol, and acetone followed by plasma cleaning for 10 minutes. Shortly after plasma cleaning, the substrate was coated with 100 μL of the perovskite solution and spun using the following process: 1000 rpm for 10 seconds followed by 5000 rpm for 60 seconds and then annealed on a hotplate at 90 °C for 60 seconds.

For the amino acid perovskites, the spin-coating solution was prepared by dissolving crystals of either D/L-alanine-PbI$_3$ or D/L-proline-PbI$_3$ in DMF to form a 1 M solution with the rest of the preparation the same as previously described. The annealing conditions for the alanine-based materials were 15 seconds at 90 °C to prevent premature decomposition of the thin films while the proline-based materials were annealed for 300 seconds at 90 °C based on literature precedent.[42]

*CD spectroscopy*

CD absorption spectroscopy was performed using a Jasco-J1500 spectrometer. The thin-films were measured at room temperature using a Xe lamp with a scanning speed of 100 nm / min and a bandwidth of 4 nm. Spectra were collected on both sides of the films and were averaged to mitigate any effects of birefringence and confirm the circular dichroism phenomenon. Cleaned, blank glass slides were used to obtain background reference spectra prior to measurement of the perovskite films.

*Circularly polarized Raman spectroscopy*

Circularly polarized Raman spectra were collected in a Renishaw inVia Raman microscope equipped with a Coherent THz Raman probe. The laser excitation for the measurements was 785 nm. The left and right circularly polarized excitation (LCP and RCP, respectively) was achieved by manually inserting a half waveplate and/or quarter waveplate. The Raman spectra were collected by focusing the 785 nm excitation laser onto the CHOIP single crystals through a 50x objective lens, with a laser power of ~ 500 μW and 10 spectral accumulations with acquisitions times set to 3 s. Temperature-dependent spectra were collected with the help of a Microptik heating/cooling stage. The crystals were loaded into the cell and evacuated to a base pressure of 100 mTorr using a mechanical vacuum pump. Low temperature spectra were collected using a recirculating liquid nitrogen pump for cooling. For these measurements a 50 x long working distance objective lens was used for focusing the excitation laser on the surface of the crystals.

*Single crystal x-ray diffraction*

The crystal structures were determined using a Rigaku XtaLAB Synergy-S, PhotonJet-I with a CCD detector using either Mo Kα ($\lambda = 0.709300$ Å) or Cu Kα radiation ($\lambda = 1.5406$ Å). Crystal samples were mounted in oil on a ring loop and placed in a cryo $N_2$ stream at 100 K. Images were interpreted and integrated with CrysAlisPRO and structures were solved using Olex2 with the ShelXT solution program and further refinement using ShelXL.[43] Non-hydrogen atoms were refined anisotropically.

CCDC 2479253-2479256 and 2480807- 2480810 contain the supplementary crystallographic data for this paper. These data are provided free of charge by the Cambridge Crystallographic Data Centre.

*Computational Details*

DFT calculations were performed with the Vienna ab initio simulation package (VASP 5.4),[44,45] applying the projector augmented-wave potential. The Kohn-Sham equations were solved using a plane wave basis set with an energy cutoff of 550 eV. The initial perovskite structures of (S-NEA)PbI$_3$ and (R-NEA)PbI$_3$ were from experiment, found to be consistent with those from Ref. [46] Structures were optimized using the Perdew-Burke-Ernzerhof (PBE) exchange-correlation functional,[47] including the D3 correction of Grimme for London dispersion,[48] unless indicated otherwise. A 4×4×1 *k*-point sampling was used. Geometries were fully relaxed regarding lattice parameters and interatomic distances until forces were less than 0.001 eV/Å. Raman spectra were calculated using Phonopy[49] and Phonopy Spectroscopy[50] at the PBE+D3 level. The Phonopy package was employed to calculate the zone-center phonon frequencies and phonon eigenvectors of the DFT optimized structures using the finite-displacement approach. The force constant matrix was constructed in a 1×1×1 supercell. The Raman spectra were simulated by appropriately averaging the Raman activity tensor calculated for each Raman-active phonon eigenmode at the zone center.

**References**

[1] W. Li, Z. Wang, F. Deschler, S. Gao, R. H. Friend, A. K. Cheetham, *Nature Reviews Materials* **2017**, *2*, 1.

[2] D. A. Egger, A. M. Rappe, L. Kronik, *Acc. Chem. Res.* **2016**, *49*, 573.

[3] J. Ma, H. Wang, D. Li, *Advanced Materials* **2021**, *33*, 2008785.

[4] Y. Xie, J. Morgenstein, B. G. Bobay, R. Song, N. A. Caturello, P. C. Sercel, V. Blum, D. B. Mitzi, *Journal of the American Chemical Society* **2023**, *145*, 17831.

[5] M. K. Jana, R. Song, H. Liu, D. R. Khanal, S. M. Janke, R. Zhao, C. Liu, Z. Valy Vardeny, V. Blum, D. B. Mitzi, *Nature communications* **2020**, *11*, 4699.

[6] X.-B. Han, W. Wang, M.-L. Jin, C.-Q. Jing, J.-M. Zhang, C.-C. Fan, *Inorganic Chemistry* **2024**, *63*, 19030.

[7] G. Long, R. Sabatini, M. I. Saidaminov, G. Lakhwani, A. Rasmita, X. Liu, E. H. Sargent, W. Gao, *Nature Reviews Materials* **2020**, *5*, 423.

[8] A. Shpatz Dayan, M. Wierzbowska, L. Etgar, *Small Structures* **2022**, *3*, 2200051.

[9] M. T. Pham, E. Amerling, T. A. Ngo, H. M. Luong, K. Hansen, H. T. Pham, T. N. Vu, H. Tran, L. Whittaker-Brooks, T. D. Nguyen, *Advanced Optical Materials* **2022**, *10*, 2101232.

[10] Y. Shao, W. Gao, H. Yan, R. Li, I. Abdelwahab, X. Chi, L. Rogée, L. Zhuang, W. Fu, S. P. Lau, *Nature communications* **2022**, *13*, 138.

[11] H. Duim, M. A. Loi, *Matter* **2021**, *4*, 3835.

[12] H. Hou, S. Tian, H. Ge, J. Chen, Y. Li, J. Tang, *Advanced Functional Materials* **2022**, *32*, 2209324.

[13] L. Wang, Y. Xue, M. Cui, Y. Huang, H. Xu, C. Qin, J. Yang, H. Dai, M. Yuan, *Angewandte Chemie* **2020**, *132*, 6504.

[14] A. Ishii, T. Miyasaka, *Science advances* **2020**, *6*, eabd3274.

[15] C. Chen, L. Gao, W. Gao, C. Ge, X. Du, Z. Li, Y. Yang, G. Niu, J. Tang, *Nature communications* **2019**, *10*, 1927.

[16] M. K. Jana, R. Song, Y. Xie, R. Zhao, P. C. Sercel, V. Blum, D. B. Mitzi, *Nature Communications* **2021**, *12*, 4982.

[17] D. Spirito, Y. Asensio, L. E. Hueso, B. Martín-García, *J. Phys. Mater.* **2022**, *5*, 034004.

[18] L.-Q. Xie, T.-Y. Zhang, L. Chen, N. Guo, Y. Wang, G.-K. Liu, J.-R. Wang, J.-Z. Zhou, J.-W. Yan, Y.-X. Zhao, *Physical Chemistry Chemical Physics* **2016**, *18*, 18112.

[19] O. Yaffe, Y. Guo, L. Z. Tan, D. A. Egger, T. Hull, C. C. Stoumpos, F. Zheng, T. F. Heinz, L. Kronik, M. G. Kanatzidis, J. S. Owen, A. M. Rappe, M. A. Pimenta, L. E. Brus, *Phys. Rev. Lett.* **2017**, *118*, 136001.

[20] M. Ledinský, P. Löper, B. Niesen, J. Holovský, S.-J. Moon, J.-H. Yum, S. De Wolf, A. Fejfar, C. Ballif, *The journal of physical chemistry letters* **2015**, *6*, 401.

[21] K. Nakada, Y. Matsumoto, Y. Shimoi, K. Yamada, Y. Furukawa, *Molecules* **2019**, *24*, 626.

[22] M. Mączka, M. Ptak, *Solids* **2022**, *3*, 111.

[23] A. Francisco-López, B. Charles, M. I. Alonso, M. Garriga, M. Campoy-Quiles, M. T. Weller, A. R. Goñi, *The Journal of Physical Chemistry C* **2020**, *124*, 3448.

[24] L. D. Barron, F. Zhu, L. Hecht, G. E. Tranter, N. W. Isaacs, *Journal of molecular structure* **2007**, *834*, 7.

[25] L. A. Nafie, *Annual Review of Physical Chemistry* **1997**, *48*, 357.

[26] E. Er, T. H. Chow, L. M. Liz-Marzán, N. A. Kotov, *ACS Nano* **2024**, *18*, 12589.

[27] T. Wang, H. Sun, X. Li, L. Zhang, *Nano Letters* **2024**, *24*, 4311.

[28] E. Oishi, Y. Fujii, A. Koreeda, *Physical Review B* **2024**, *109*, 104306.

[29] K. Ishito, H. Mao, K. Kobayashi, Y. Kousaka, Y. Togawa, H. Kusunose, J. Kishine, T. Satoh, *Chirality* **2023**, *35*, 338.

[30] K. Ishito, H. Mao, Y. Kousaka, Y. Togawa, S. Iwasaki, T. Zhang, S. Murakami, J. Kishine, T. Satoh, *Nat. Phys.* **2023**, *19*, 35.

[31] S. Zhang, N. Mao, N. Zhang, J. Wu, L. Tong, J. Zhang, *ACS nano* **2017**, *11*, 10366.

[32] S. Zhang, J. Huang, Y. Yu, S. Wang, T. Yang, Z. Zhang, L. Tong, J. Zhang, *Nature Communications* **2022**, *13*, 1254.

[33] S. Yang, Y. Yu, F. Sui, R. Ge, R. Jin, B. Liu, Y. Chen, R. Qi, F. Yue, *ACS Nano* **2024**, *18*, 33754.

[34] R. Rao, J. Jiang, R. Pachter, T. T. Mai, V. Mohaugen, M. F. Muñoz, R. Siebenaller, E. Rowe, R. Selhorst, A. N. Giordano, *ACS Nano* **2025**, *19*, 26377.

[35] M. Pols, G. Brocks, S. Calero, S. Tao, **2024**, DOI: 10.48550/arXiv.2411.17225.

[36] G. Li, M. Alshalalfeh, J. Kapitán, P. Bouř, Y. Xu, *Chemistry–A European Journal* **2022**, *28*, e202104302.

[37] L. Barron, A. Buckingham, *Molecular Physics* **1971**, *20*, 1111.

[38] L. Hecht, L. Barron, A. Gargaro, Z. Wen, W. Hug, *Journal of Raman spectroscopy* **1992**, *23*, 401.

[39] L. Zhang, Q. Niu, *Physical Review Letters* **2014**, *112*, 085503.

[40] L. Zhang, *Phys. Rev. Lett.* **2015**, *115*, DOI: 10.1103/PhysRevLett.115.115502.

[41] M. Hamada, E. Minamitani, M. Hirayama, S. Murakami, *Phys. Rev. Lett.* **2018**, *121*, 175301.

[42] M. Xin, P. Cheng, X. Han, R. Shi, Y. Zheng, J. Guan, H. Chen, C. Wang, Y. Liu, J. Xu, *Advanced Optical Materials* **2023**, *11*, 2202700.

[43] O. V. Dolomanov, L. J. Bourhis, R. J. Gildea, J. A. Howard, H. Puschmann, *Applied Crystallography* **2009**, *42*, 339.

[44] G. Kresse, J. Furthmüller, *Computational materials science* **1996**, *6*, 15.

[45] G. Kresse, D. Joubert, *Phys. Rev. B* **1999**, *59*, 1758.

[46] L.-L. Zhu, Y.-E. Huang, Y.-P. Lin, X.-Y. Huang, H.-Q. Liu, D. B. Mitzi, K.-Z. Du, *Polyhedron* **2019**, *158*, 445.

[47] J. P. Perdew, K. Burke, M. Ernzerhof, *Physical review letters* **1996**, *77*, 3865.

[48] S. Grimme, J. Antony, S. Ehrlich, H. Krieg, *J. Chem. Phys.* **2010**, *132*, 154104.

[49] A. Togo, I. Tanaka, *Scripta Materialia* **2015**, *108*, 1.

[50] J. M. Skelton, L. A. Burton, A. J. Jackson, F. Oba, S. C. Parker, A. Walsh, *Phys. Chem. Chem. Phys.* **2017**, *19*, 12452.